\documentclass{aastex}
\usepackage{emulateapj5}
\usepackage{onecolfloat5}
\input epsf

\def\plotone#1{\centering \leavevmode

\epsfxsize= 1.0\columnwidth \epsfbox{#1}}

\def\C{ {\cal C}}

\makeatletter

\newenvironment{tablehere}{\def\@captype{table}}{}

\makeatother

\long\def\comment#1{}

\def\W2{{\cal W}}

\newcommand{\tableskip}{\\[-6pt]}

\def\be{\begin{equation}}
\def\ee{\end{equation}}
\def\bea{\begin{eqnarray}}
\def\eea{\end{eqnarray}}

\def\cmm2{{\,\rm cm^{-2}}}

\def\cm2{{\,{\rm cm}^2}}

\def\cmm3{{\,{\rm cm}^{-3}}}

\def\gcmm3{{\,{\rm g\,cm^{-3}}}}

\def\HO{{100h\,{\rm km\,sec^{-1}\,Mpc^{-1}}}}

\def\fun#1#2{\lower3.6pt\vbox{\baselineskip0pt\lineskip.9pt

  \ialign{$\mathsurround=0pt#1\hfil##\hfil$\crcr#2\crcr\sim\crcr}}}

\lefthead{KNOX ET AL.}

\righthead{}

\begin{document}
\twocolumn[
\submitted{Submitted to ApJ Letters}
\title{The Age of the Universe and the Cosmological Constant Determined
from Cosmic Microwave Background Anisotropy Measurements}
\author{Lloyd Knox}
\affil{Department of Physics, University of California, Davis, CA
95616, USA, email: lknox@ucdavis.edu}
\author{Nelson Christensen} \affil{Physics and Astronomy, Carleton
College, Northfield, MN, 55057, USA email: nchriste@carleton.edu}
\and
\author{Constantinos Skordis} \affil{Department of Physics,
University of California, Davis, CA 95616, USA, email:
skordis@bubba.ucdavis.edu}
\begin{abstract}
If $\Omega_{\rm tot} = 1$ and structure formed from adiabatic
initial conditions then the age of the Universe, as constrained by
measurements of the cosmic microwave background (CMB), is
$t_0=14.0 \pm 0.5$ Gyr.  The uncertainty is surprisingly small
given that CMB data alone constrain neither $h$ nor
$\Omega_\Lambda$ significantly.  It is due to the tight (and
accidental) correlation, in these models, of the age with the
angle subtended by the sound horizon on the last--scattering
surface and thus with the well-determined acoustic peak locations.
If we assume either the HST Key Project result $h = 0.72 \pm .08$
or simply that $h > 0.55$, we find $\Omega_\Lambda > 0.4$ at 95\%
confidence---another argument for dark energy, independent of
supernovae observations. Our analysis is greatly simplified by the
Monte Carlo Markov chain approach to Bayesian inference combined
with a fast method for calculating angular power spectra.
\end{abstract}
 ]
\section{Introduction}

Determining the expansion age of the Universe has been a major
goal of cosmology ever since Hubble discovered the expansion.
Compatibility with determinations of stellar ages is an important
consistency check of cosmological models.  Traditional methods of
determining the expansion age rely on Hubble constant measurements
which are either highly imprecise, or have error budgets dominated
by systematics.  In addition one must determine $\Omega_\Lambda$
(or more generally the mean density of the various components)
since it affects how the expansion rate has changed over time. In
this {\it Letter} we present highly precise age determinations
from CMB data which completely bypass the need for independent
determinations of $H_0$ and $\Omega_\Lambda$.

There are a handful of cosmological parameters which can be
determined from measurements of the CMB angular power spectrum to
percent level accuracy, such as $\omega_b$, $\omega_m$, and
$\Omega_{\rm tot}$ (where $\omega_i \equiv \Omega_i h^2$ and $H_0
= \HO$) \citep[e.g.][]{eisenste99}. Other parameters can not be
well--determined and require the addition of complementary
observations.  For example $\Omega_\Lambda$ is poorly determined
by the CMB alone\citep[e.g.][]{efstathiou99} but well--determined
when supernovae observations are included
\citep[e.g.][]{netterfield01}.

It has been pointed out \citep{ferreras01} and demonstrated 
\citep{netterfield01} that the CMB can be used to place
tight constraints on the age of the Universe.  This is due to the high
degree of correlation between the angle subtended by the
sound--horizon on the last--scattering surface, $\theta_s$, and age in
flat adiabatic models, also noticed by \citet{hu01} who used it to
place an {\it upper} bound on the age.  Here we extend the previous
work by including additional data, 
by taking the flatness
assumption seriously and by use of a new analysis technique which has
advantages as described below.  We also demonstrate the accidental
nature of the age--sound horizon correlation by showing that its
tightness depends on where we are in the $\Omega_\Lambda, h$ parameter
space.  We are fortunate that the correlation is tightest near the
``concordance'' values of $h=0.72$ and $\Omega_\Lambda = 0.65$.

Our age determination is model--dependent and the model (adiabatic
CDM) has many parameters.  We take them to be the amplitude and
power--law spectral index of the primordial matter power spectrum,
$A$ and $n$, the baryon and dark matter densities, $\omega_b$ and
$\omega_d$, the cosmological constant divided by the critical
density, $\Omega_\Lambda$, and the redshift of re-ionization of
the intergalactic medium, $z_{\rm ri}$.  The Hubble constant and
age are derived parameters, given in terms of the others by $h^2 =
(\omega_b + \omega_d)/(1-\Omega_\Lambda)$ and $t_0 = 6.52 \ {\rm
Gyr} \
\ln[(1+\sqrt{\Omega_\Lambda})/\sqrt{1-\Omega_\Lambda}]/\sqrt{\Omega_\Lambda
h^2}$.

We do not consider models with $\Omega_{\rm tot} \ne 1$ or dark
energy models other than the limiting case of a cosmological
constant with $w \equiv P/\rho= -1$. The first we justify on
grounds of simplicity:  CMB observations indicate the mean
curvature is close to zero and generally agree well with
inflation. If we did allow the curvature to vary, our age result
would become significantly less precise. We expect allowing $w$ to
vary to have little effect, as we discuss below.

We explore the likelihood in a ten--dimensional parameter space
(six cosmological parameters plus four experimental parameters) by
Monte--Carlo generation of a Markov chain of parameter values as
described in \citet{christensen01}. From the chain one can rapidly
calculate marginalized one--dimensional or two--dimensional
probability distributions for chain parameters, or derived
parameters, with or without additional priors.
Generating a sufficiently long chain in a reasonable amount of
time requires a fast means of calculating the angular power
spectrum for a given model. We describe this fast method briefly
below and more thoroughly in \citet{kaplinghat01}.

Supernovae observations constrain the combination $H_0 t_0$ better
than either parameter by itself.  
\citet{perlmutter99a} find for flat Universes that $t_0=
13.0^{+1.2}_{-1.0} (0.72/h)$ Gyr. 
Combining this result with our $t_0$ determination leads to
$h = 0.67^{+.07}_{-.06}$, in agreement with the HST result.
\citet{riess98} find for
arbitrary $\Omega_{\rm tot}$ that $t_0= (14.2 \pm 1.7)$ Gyr.

\citet{krauss01} estimate the age of 17 metal--poor globular
clusters to be $t_{\rm GC}$ = 12.5 Gyr with a 95\% lower bound of
$10.5$ Gyr and a 68\% upper bound of 14.4 Gyr.  The minimum
requirement for consistency, that $t_f \equiv t_0 - t_{\rm GC} >
0$ is easily satisfied with a few Gyrs to spare. Unfortunately,
the upper bound on $t_{\rm GC}$ is not sufficiently restrictive to
set an interesting lower bound on $t_f$.

Below we tabulate our constraints on all the model parameters, and
emphasize not only $t_0$ but also $\Omega_\Lambda$. With the
inclusion of prior information on $H_0$, the CMB data provide
strong evidence for $\Omega_\Lambda > 0$.  The same conclusion can
be reached by, instead, combining the CMB data with observations
of large--scale structure \citep{efstathiou01a} or clusters of
galaxies \citep{dodelson00}.

\section{Method}

Our first step in exploring the high--dimensional parameter space
is the creation of an array of parameter values called a chain,
where each element of the array, $\vec \theta$, is a location in
the $n$-dimensional parameter space.  The chain has the useful
property once it has converged that $P(\vec \theta \in R) =
N(\vec \theta \in R)/N$  where the left--hand side is the
posterior probability that $\vec \theta$ is in the region $R$, $N$
is the total number of chain elements and $N(\vec \theta\in R)$ is
the number of chain elements with $\vec \theta$ in the region $R$.
Once the chain is generated one can then rapidly explore
one--dimensional or two--dimensional marginalizations in either
the original parameters, or in derived parameters, such as $t_0$.
Calculating the marginalized posterior distributions is simply a
matter of histogramming the chain.

\subsection{Generating the Chain}

The chain we generate is a Monte Carlo Markov Chain (MCMC) produced
via the Metropolis--Hastings algorithm described in Christensen et
al. (2001).  The candidate--generating function for an initial run was
a normal distribution for each parameter.  Subsequent runs
used a multivariate--normal distribution with cross--correlations
between cosmological parameters equal to those of the posterior as
calculated from the initial run.

All of our results are based on MCMC runs consisting of $2 \times
10^5$ iterations. For the ``burn-in'' the initial $2.5 \times 10^4$
samples were discarded, and the remaining set was thinned by accepting
every 25th iteration.  We used the CODA software \citep{best95} to
confirm that all chains passed the Referty-Lewis convergence
diagnostics and the Heidelberger-Welch stationarity test.

While generating the chain we always restrict our sampling to the
$h > 0.4$ and $5.8 < z_{\rm ri} < 6.3$ region of parameter space.
The former is a very conservative lower--bound on $h$ and the
latter is a simple interpretation of the spectra of quasars at
very high redshift \citep{becker01,djorgovski01}.  For some of our
results we assume an ``HST prior'' which means $h=0.72 \pm 0.08$
\citep{freedman01} with a normal distribution.

\subsection{$C_l$ Calculation}

We calculate $C_l$ rapidly with a preliminary version of
the Davis Anisotropy Shortcut \citep[DASh;][]{kaplinghat01}.
We first calculate the Fourier and
Legendre--transformed photon temperature perturbation,
$\Delta_l(k)$, on a grid over parameters $\omega_b$ and $\omega_d$
at fixed values of $\Omega_k \equiv 1-\Omega_{\rm tot}=
\Omega_k^*$, $\Omega_\Lambda = \Omega_\Lambda^*$ and $\tau=0$
using CMBfast \citep{seljak96}. From this grid, we get $C_l$ for
any $\omega_b$, $\omega_d$, and the primordial power spectrum
$P(k)=A(k/0.05 Mpc^{-1})^n$ by performing multi--linear
interpolation on the grid of $\Delta_l(k)$ and then the following
integral: \be \label{eqn:clint} {\cal C}_l \equiv {l(l+1)C_l\over 2\pi} = 
8\pi l (l+1)  \int k^2 dk
\Delta_l^2(k) P(k). \ee We can get any $C_l$ in the entire model
space of \{$\omega_b$, $\omega_d$, $\tau$, $\Omega_\Lambda$,
$\Omega_k$, $P(k)$\} by the use of analytic relations between the
$\Delta_l(k)$ for different models.  For varying $\Omega_\Lambda$
and $\Omega_k$, ${\cal C}_l = {\cal C}_{\tilde l}$ where $l/\tilde
l =
\theta_s(\Omega_k^*,\Omega_\Lambda^*)/\theta_s(\Omega_k,\Omega_\Lambda)$
and $\theta_s$ is the angle subtended by the sound--horizon at the
last--scattering surface.  For $\Omega_k=0$, $\theta_s = s/\eta_0$
where $\eta_0$ is the conformal time today (or, equivalently, the
comoving distance to the horizon) and $s$ is the comoving
sound--horizon at the last--scattering surface.

Altering $\theta_s$ is not the only effect of varying $\Omega_k$
and $\Omega_\Lambda$.  Varying $\Omega_k$ changes the eigenvalues
of the Laplacian on very large scales and hence the power spectrum
at the last--scattering surface, and both $\Omega_k$ and
$\Omega_\Lambda$ affect the late--time evolution of the
gravitational potential. Both of these effects only affect $C_l$
at $l << 100$.  We therefore make an additional grid over the
parameters $\omega_b$, $\omega_d$, $\Omega_k$ and $\Omega_\Lambda$
but with smaller maximum $l$ and $k$ values than the
lower--dimensional high-$l$ grid. The low-$l$ grid used for the
calculations presented here has ranges $0.01 < \omega_b < 0.03$,
$0.05 < \omega_d < 0.25$, and $0 < \Omega_\Lambda <0.85$ with 4,
4, and 8 uniformly spaced samplings of the range respectively. For
the present application we have fixed $\Omega_k=0$.  For the
high-$l$ grid the ranges for $\omega_b$ and $\omega_d$ are the
same but with twice as many samples, and $\Omega_\Lambda^* = 0.6$.

The split into a low-$l$ grid and a high-$l$ grid has been used by
others, although for grids of $C_l$, not $\Delta_l(k)$.  We follow
\citet{tegmark01} in joining our grids with a smooth k-space
kernel, $g(k)= 2/(1+exp(2k/k_s)^4)$ where $k_s = 1.5/s$; in the
integrand of Eq.~\ref{eqn:clint} $P(k)$ is replaced with
$g(k)P(k)$ for the low-$l$ grid and $(1-g(k))P(k)$ for the
high-$l$ grid and $C_l = C_l^{\rm low} + C_l^{\rm high}$. Finally,
we allow for non-zero $z_{ri}$ by sending $C_l \rightarrow {\cal
R}_l(z_{ri})C_l$ where ${\cal R}_l(z_{ri})$ is given by the
fitting formula of \citet{hu97}.

\subsection{Likelihood Calculation}

To calculate the likelihood we use the offset log-normal approximation
of \citet{bjk00} which is a better approximation to the likelihood
function than a normal distribution.  We include bandpower data from
Boomerang, the Degree Angular Scale Interferometer
\citep[DASI;][]{halverson01}, Maxima \citep{lee01}) and the COsmic
Background Explorer\citep[COBE;][]{bennet96}.  The weight matrices,
band powers and window functions for DASI are available in
\citet{leitch01}, \citet{halverson01} and \citet{pryke01}. For COBE we
approximate the window functions as tophat bands; all other
information is available in \citet{bjk00} and in electronic form at
\citet{radpack}.  For Boomerang and Maxima we approximate the window
functions as top-hat bands, the weight matrices as diagonal and the
log-normal offsets, $x$, as zero. The Boomerang team report the
uncertainty in their beam full-width at half-maximum (fwhm) as $12.9
\pm 1.4$ minutes of arc. We follow them in modelling the departure
from the nominal (non-Gaussian) beam shape as a Gaussian. For the
Boomerang $Z_i^t$, $\C_l$ is therefore actually $\C_l \exp(-l^2
b^2)$. Our prior for $b$ is uniform, bounded such that the fwhm is
always between 11.5' and 14.3'.  Calibration parameters, for example,
$u_{\rm DASI}=1 \pm 0.04$ are taken to have normal prior distributions
and alter model angular power spectra via $C_l \rightarrow u_{\rm
DASI}^2 C_l$ prior to comparison with the reported band powers.  To
reduce our sensitivity to beam errors, we only use bands with maximum
$l$-values less than 1000. Thus we use all nine DASI bands, the first
11 Maxima bands and all but the last Boomerang band.

Five of the 24 DASI fields are completely within the area of sky
analyzed by Boomerang, and three partially overlap this area.  We
expect the resulting DASI--Boomerang bandpower error correlations (which
we neglect) to be small and to have negligible effect on our results.
\section{Results}

\begin{figure}
\plotone{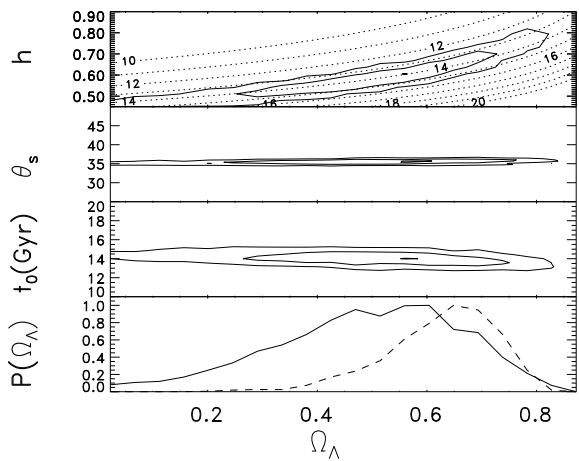}
\caption[]{\label{fig:fig1} The posterior probability density 
of $\Omega_\Lambda$
(lowest panel) and contours of equal probability density in the
$\Omega_\Lambda$, $t_0$ plane (lower middle panel),
$\Omega_\Lambda$, $\theta_s$ plane (upper middle panel) and
$\Omega_\Lambda$, $h$ plane (top panel).  Contour levels are at
$e^{-6.17/2}$, $e^{-2.3/2}$ and 0.95 of maximum. The lowest panel
curves are for $h > 0.4$ (solid) and $h = .72 \pm .08$ (dashed).
Top panel dotted lines are at constant $t_0$.}
\end{figure}

\begin{figure}
\plotone{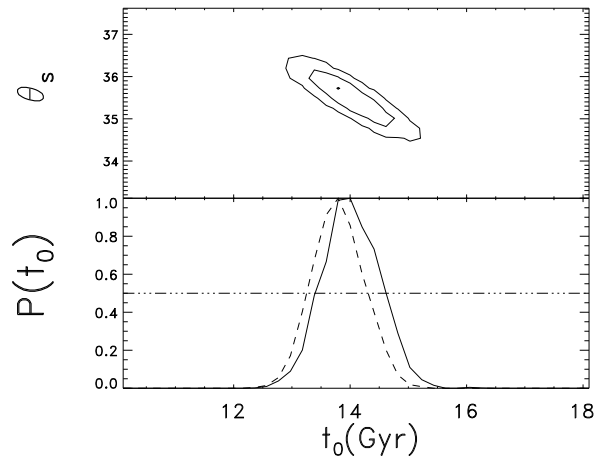}
\caption[]{\label{fig:fig2} The posterior probability density of the age of the
Universe (lower panel) and contours of equal probability density (as in
Figure 1) in the $\theta_s$, $t_0$ plane (upper panel). The lower
panel curves are for $h > 0.4$ (solid) and $h = .72 \pm .08$.  }
\end{figure}

In the top panel of Figure ~\ref{fig:fig1} we see that
neither $h$
nor $\Omega_\Lambda$ can be constrained well by CMB data alone.
The shape of the contour
``banana'' in the top panel is determined by the 
dependence of the well--determined $\theta_s$ on $h$ and $\Omega_\Lambda$;  
lines of constant $\theta_s$ run along the ridge of high likelihood.
Since $\theta_s$ correlates well with the age, as seen in Fig.~\ref{fig:fig2},
the ridge of high likelihood is also at nearly constant age.

Although $\Omega_\Lambda$ is poorly determined by CMB data alone
we find that addition of the HST prior allows one to set a 95\%
lower limit of $\Omega_\Lambda > 0.4$.  This same lower bound can
be achieved by simply rejecting models with $h < 0.55$.  

We can understand the $t_0--\theta_s$ correlation 
with the aid of an approximate analytic
expression given by \citet{hu01} from which we derive \be
\label{eqn:lA} {\Delta \theta_s\over \theta_s} = 0.060(2.9{\Delta
\omega_m \over \omega_m}+1.0{\Delta \omega_\Lambda \over
\omega_\Lambda}-1.14{\Delta \omega_b \over \omega_b}) \ee for the
fiducial values $\omega_m=0.15, \omega_\Lambda = 0.3,
\omega_b=0.02$. Expanding $t_0$ about our fiducial values we find
\be {\Delta t_0 \over t_0} = -0.12\left(3.0{\Delta \omega_m \over
\omega_m}+1.2{\Delta \omega_\Lambda \over
\omega_\Lambda}\right).\ee Thus a change from the fiducial values
by $\Delta \omega_m$ and $\Delta \omega_\Lambda$ which keeps
$\theta_s$ fixed will nearly leave the age unchanged.

In general, the parameter controlling this correlation is the
ratio of ratios: \be R = { (\partial \ln \theta_s /\partial \ln
\omega_m) / (\partial \ln \theta_s /\partial \ln \omega_\Lambda)
\over(\partial \ln t_0 /\partial \ln \omega_m) / (\partial \ln t_0
/\partial \ln \omega_\Lambda) } \ee and the correlation is
tightest when $R=1$.  $R$ has little dependence on $\omega_b$.
For what used to be called standard CDM, $R = 0.75$. R can be as
small as 0.53 for $\Omega_m=1$ and $h=0.72$ and as large as 2.0
for $\Omega_\Lambda=0.83$ and $h=0.72$.  At the maximum of the
likelihood, $R=0.88$(1.08) with (without) the HST prior.

As a test, we have estimated the age via a direct grid--based
evaluation of the likelihood given DASI and DMR data 
using CAMB \citep{camb} to calculate $C_l$'s.
Taking the grid parameters to be $t_0$, $\omega_d$, $n$ and $A$ 
and fixing
$u_{\rm DASI}-1 = \omega_b-1 = \Omega_k =
z_{\rm ri} = 0$ we find
$t_0=(13.6
\pm 0.6)$ Gyr.  This agrees very well with our MCMC + DASh results
when the same assumptions and data selection are made: $t_0 =
(13.7 \pm 0.6)$ Gyr.  We can also reproduce the \citet{netterfield01}
result, finding for Boomerang and DMR data (although ignoring
the highest--$l$ Boomerang bandpower) $t_0 =
(14.32 \pm 0.68$) Gyr.

In the table we show means and standard deviations for our ten
original parameters and a number of derived parameters.
Particularly noteworthy is $\theta_s$, determined with an error of
less than 3\%.  The agreement between the datasets on this number
is also remarkable.  From DASI $\theta_s = (0.60 \pm 0.01)$
degrees and from Boomerang $\theta_s = (0.59 \pm 0.01)$ degrees.

For studying the early evolution of structure, it is useful to
know the age at redshifts in the matter--dominated era. For $1 <<
z <100$, $t_z \times (1+z)^{3/2} = 6.52/\sqrt{\omega_m}$ Gyr =
$(16 \pm 1)$ Gyr, where $t_z$ is the age at redshift $z$. Since
{\it MAP}\footnote{{\it MAP}:  http://map.gsfc.nasa.gov} and {\it
Planck}\footnote{{\it Planck}: http://astro.estec.esa.nl/Planck/}
will determine $\omega_m$ to 10\% and 2\% respectively
\citep{eisenste99} they will determine $t_z \times (1+z)^{3/2}$ to
5\% and 1\% respectively.

\begin{tablehere}
\begin{table*}[hbt]\small
\caption{\label{table:bounds}}
\begin{center}
{\sc Parameter Bounds}\\
\begin{tabular}{ccc}
\tableskip\hline\hline\tableskip Parameter & mean & standard deviation \\
\tableskip\hline\tableskip
               $\omega_b$&   0.021&   0.002\\
               $\omega_d$&   0.145&   0.021\\
         $\Omega_\Lambda^1$&   0.49&   0.17\\
             $z_{\rm ri}^1$&   6.0&   0.14\\
                      $A$&   6.7&   0.56\\
                      $n$&   0.96&   0.04\\
           $u_{\rm DASI}$&   1.00&   0.03\\
           $u_{\rm Boom}$&   1.07&   0.03\\
         $u_{\rm Maxima}$&   1.00&   0.03\\
           fwhm$_{\rm Boom}^1$&  13'.9&   0'.3\\
\hline
              $t_0$ (Gyr)&  14.0&   0.48\\
                      $h^1$&   0.59&   0.07\\
               $\Omega_m^1$&   0.51&   0.17\\
        $\omega_\Lambda^1 $&   0.19&   0.11\\
               $\omega_m$&   0.166&   0.021\\
         $c \eta_0$ (Gpc)&  13.95&   0.56\\
     $\theta_s$ &   35'.5&   0'.43\\
             $l_{\rm eq}$& 168&  15\\
                    $l_d$&1392&  18\\
                    $H_2$&   0.481&   0.024\\
                    $H_3$&   0.486&   0.030\\
 $t_z \times (1+z)^{3/2}$&  16.0&   1.0\\

\tableskip\hline
\end{tabular}\\[12pt]
\begin{minipage}{5.2in}
NOTES.---%
The mean and standard deviations for the ten chain parameters
(top) plus derived parameters (bottom).  For these results we use
all the data with our weakest prior assumptions.  Units of $A$ are
arbitrary. See \citet{hu01} for the definition of $l_{\rm eq}$,
$l_d$, $H_2$ and $H_3$.  The $1$ superscript indicates those
parameters whose uncertainties are not well--described by a mean
and standard deviation. For example, the posterior probability
distribution for $z_{\rm ri}$ is not significantly different from
the prior one we assumed, which is uniform between 5.8 and 6.3.
\end{minipage}
\end{center}
\end{table*}
\end{tablehere}
\section{Discussion}

Since our argument is model--dependent, it is worth pointing out
that the model has been enormously successful on the relevant
length scales \citep[e.g.][]{wang01}. Perhaps the weakest link is
the dark energy equation of state since we have scant guidance
from observations \citep[e.g.][]{perlmutter99b} and even less from
theory. Fortunately, one can show that varying the equation of
state away from $w=-1$ at fixed $\theta_s$ has very little effect
on the age: at $\theta_s = 0.6$ degrees and $\omega_m = 1/3$,
$\Delta t_0/t_0 = .05 \Delta w$. Though we neglect the possibility
of gravitational wave contributions to $C_l$
\citep[see][]{efstathiou01b} we do not expect these to make much
difference since $\theta_s$ is relatively unaffected by
measurements at low $\ell$ where gravitational waves are
important.

Our age determination has the benefit of being derived from
observations whose statistical properties can be predicted highly
accurately using linear perturbation theory.  We are encouraged
that the observational errors are dominated by the reported
statistical ones since nearly the same result can be derived from
two independent data sets.  We conclude that the best
determination of $t_0$ now comes from CMB data.  The prospects for
improving the age determination are bright since the statistical
errors (and any systematic ones too) will be greatly reduced by
{\it MAP} data in the near future.

The MCMC chains we have generated are available via e-mail from
the authors.

 \acknowledgments We thank M. Kaplinghat, R. Meyer
and K. Ganga for useful conversations, B. Luey for some programming, B.
Chaboyer for sharing results prior to publication and the Fermilab
Reading Group for comments on an earlier version. LK is supported
by NASA, NC by the NSF and CS by the DoE.

\end{document}